# First-principles study of electron and hole doping in perovskite nickelates

Lucia Iglesias [1], Manuel Bibes[1] and Julien Varignon[2]

[1] Unité Mixte de Physique, CNRS, Thales, Université Paris Sud, Université Paris-Saclay, F-91767 Palaiseau, France
[2] Laboratoire CRISMAT, CNRS UMR 6508, ENSICAEN, Normandie Université, 6 boulevard Maréchal Juin, F-14050 Caen Cedex 4, France

## Abstract

Rare-earth nickelates $R^{3+}Ni^{3+}O_3$ (R=Lu-Pr, Y) show a striking metal-insulator transition in their bulk phase whose temperature can be tuned by the rare-earth radius. These compounds are also the parent phases of the newly identified infinite layer $RNiO_2$ superconductors. Although intensive theoretical works have been devoted to understand the origin of the metal-insulator transition in the bulk, there have only been a few studies on the role of hole and electron doping by rare-earth substitutions in $RNiO_3$ materials. Using *first-principles* calculations based on density functional theory (DFT) we study the effect of hole and electron doping in a prototypical nickelate $SmNiO_3$. We perform calculations without Hubbard-like U potential on Ni *3d* levels but with a meta-GGA better amending self-interaction errors. We find that at low doping, polarons form with intermediate localized states in the band gap resulting in a semiconducting behavior. At larger doping, the intermediate states spread more and more in the band gap until they merge either with the valence (hole doping) or the conduction (electron doping) band, ultimately resulting in a metallic state at 25% of R cation substitution. These results are reminiscent of experimental data available in the literature and demonstrate that DFT simulations without any empirical parameter are qualified for studying doping effects in correlated oxides and to explore the mechanisms underlying the superconducting phase of rare-earth nickelates.

## I. Introduction

Transition metal oxide perovskites $ABO_3$, with a *3d* element sitting on the B site, are an important class of materials that shows a vast array of functionalities such as ferroelectricity, magnetism or superconductivity for instance[1]. Among these perovskites, rare-earth nickelates $R^{3+}Ni^{3+}O_3$ (R=Lu-Pr, Y) have attracted a lot of attention for several reasons[2,3]: (i) they exhibit a tunable metal-insulator transition as a function of the rare-earth size in their bulk form, (ii) they were proposed as magnetically induced ferroelectric compounds[4] and (iii) due to their proximity with Cu in the periodic table, they have been suggested to potentially show cuprate-like superconductivity[5]. The triggering mechanism of the metal-insulator transition and the precise electronic structure of the insulating phase have been established recently[6–11]. They rely on an electronic instability associated with an unstable 3+ formal oxidation state on Ni cations that leads to charge and bond disproportionation: Ni sites split into apparent $Ni^{2+}$ and $Ni^{4+}$ cations, resulting in an alternative expansion/contraction of $O_6$ octahedra, the so-called breathing distortion (see Figure 2 c of Ref.[8]). Although ligand holes exist in nickelates due to their negative charge transfer nature[8,12,13], cuprate-like superconductivity in these compounds remained a dream for more than 30 years[5]. It is only in 2019 that superconductivity was finally achieved in nickelates through an appropriate chemical reduction to an infinite layer phase $RNiO_2$ and A site cation substitution resulting in hole doping[14,15]. Although the superconducting temperature (~15 K) remains relatively low with respect to that of the cuprates, this achievement offers a new and alternative playground to test our solid state theories of pairing mechanisms in correlated oxide superconductors[16].

The $RNiO_2$ phase has been widely studied theoretically since the discovery of superconductivity in 2019. However, there are actually very few studies on doping effects in the $RNiO_3$ perovskite phase, even though this phase is the starting point to achieve the infinite-layer superconducting phase. Electron doping effects through hydrogen insertion or oxygen vacancies have been studied[17–22], altering the metal-insulator transition and band gap amplitude, but the A site cation substitution has been less studied. Hole and electron doping through Ca or Sr and Th or Ce substitutions, respectively, have been studied experimentally in bulk and thin films of $NdNiO_3$ and $SmNiO_3$, revealing a decrease of the metal-insulator transition temperature for substitutions as large as 10%[23–25] or a metallic state at doping levels of 30%[26]. At the theoretical level, electronic calculations on trends in electronic and structural properties with substitution of A site cations are missing in the $RNiO_3$ family. Understanding polaron formation, the evolution of band gaps and lattice distortions upon doping in the $ABO_3$ phase is crucial for several reasons : (i) it may provide potential insights in the underlying superconducting $RNiO_2$ phase and/or (ii) we may benchmark our electronic structure calculations performed with DFT-no-U for studying doping effects in the parent $RNiO_3$ phase, to then transpose it to the reduced superconducting $RNiO_2$ phase. In this regard, previous studies shown that Density Functional Theory (DFT) calculations involving the meta-GGA SCAN[27] functional can successfully address the physics of bulk *3d* transition metal $ABO_3$ compounds[28] without any empirical U parameter or the doping effects in the cuprates superconductor $La_2CuO_4$[29,30].

In this manuscript, we inspect the role of electron and hole doping through Sm substitution in $SmNiO_3$ with *first-principles* calculations based on DFT without empirical U parameter. We find that at low hole or electron doping content, intermediate acceptor or donor states form in the band gap, producing a semi-conducting state. Upon increasing the doping content, we observe that intermediate states spread more and more in the band gap until they merge with the valence or conduction band, resulting in a strong decrease of the band gap. This is further accompanied by a vanishing of the breathing mode distortion, hinting at the appearance of the metallic phase. Our DFT-no-U simulations, performed with the recent meta-GGA SCAN functional better amending self-interaction errors inherent to DFT, provide results consistent with more classical DFT+U approach and with experimental results available in the literature. This validates the use of the SCAN functional for studying doping effects in $ABO_3$ materials and the underlying mechanisms yielding superconductivity in correlated oxides.

## II. Method

*The technique*: *First-principles* calculations are performed with Density Functional Theory (DFT) using the Vienna Ab initio Simulation Package (VASP)[31,32] . While it is often believed that more complex techniques than DFT (e.g. Dynamical Mean Field Theory) are required to describe the physics of correlated oxides, we recently showed that, even without any empirical U potential on Ni *3d* levels, DFT can provide a fine description of the physics of transition metal oxide $ABO_3$ perovskites[7,28] if one properly includes all degrees of freedom (symmetry lowering events, charge and spin orders, different local motifs) in the simulations.

One must also involve an exchange-correlation functional properly correcting self-interaction errors inherent to DFT, thus it is necessary to go beyond standard Local Density approximation (LDA) or Generalized Gradient Approximation (GGA) functionals. Two exchange-correlation functionals are used in this work: (i) the meta-GGA SCAN functional with no U parameter[27] and (ii) the PBEsol[33] functional along with a U potential of 2 eV on Ni $d$ states in order to better cancel self-interactions errors, previously shown to be suited for SmNiO$_3$[8], and used as a benchmark for the SCAN functional. Most notably, both functionals have revealed the electronic structure in the insulating phase and enabled a deep understanding of the mechanism producing the metal-insulator phase transition[7,8,28,34]. In RNiO$_3$ materials, the band gap is strongly sensitive to the magnetic order[8]. Since the low-temperature phase usually inherits the physics of the high temperature paramagnetic state[7], we used the complex AFM-S order appearing in the low temperature phase of bulk SmNiO$_3$. It consist of up-up-down-down spin chains in the (ab)-plane with different stacking along the c axis[4]. No exploration of magnetic orders has been performed and it is left for possible future studies. The monoclinic P2$_1$/n structure corresponding to a ($2\sqrt{2}$a, $\sqrt{2}$a, 4a) unit cell (16 formula unit, a is the primitive cubic cell lattice parameter) allowing for the complex AFM-S order as well as disproportionation effects is used for all simulations. The cut-off energy is set to 500 eV along with a 3×6×2 Gamma centered K mesh. Projected Augmented Wave (PAW) potentials[35] without treating explicitly *4f* electrons for Sm were used in the simulations. Geometry optimizations are performed until forces acting on each atom are lower than 0.05 eV/Å. Finally, the amplitude of lattice distortions appearing in the identified ground states are extracted using symmetry mode analysis with ISOTROPY applications[36,37].

*The method to describe doping effects*: A site cation substitution is modelled within the ($2\sqrt{2}$a, $\sqrt{2}$a, 4a) unit cell (16 formula unit with respect to the primitive cubic cell) with atomic position obtained using the SQS method[38], which allows to extract within a given supercell the configuration maximizing the disorder characteristic of an alloy. Hole (electron) doping is achieved by replacing Sm$^{3+}$ with Ca$^{2+}$ (hole doping) or Ce$^{4+}$ (electron doping). Proper simulation of doping effects requires to study polaron formation in the material. To that end, we have used the modified VASP routine to initially force orbital occupancy of transition metal $d$ states with DFT+U[39]. Starting from the pristine material, we force the hole(s) or the electron(s) added to the system to localize on a precise Ni site in the vicinity of the substituted cations by fixing the $e_g$ level occupancies. Then, we performed few steps of structural relaxations without preserving the symmetry of the wavefunction before switching off the constraint and let the solver relax to the ground state with either the PBEsol+U or SCAN-no-U functional.

### III. Results

#### A. *SmNiO$_3$ ground state electronic properties*

We have first relaxed SmNiO$_3$ bulk material with both SCAN and PBEsol+U functionals. Consistently with previous DFT studies[6–8], the monoclinic cell is stable in our DFT simulations showing two Ni sublattices: a Ni$_L$ cation sits in large O$_6$ octahedra and bears a magnetic moment larger than one ($\mu_{NiL}$=1.34 $\mu_B$ and $\mu_{NiL}$=1.21 $\mu_B$ with SCAN and PBEsol+U functionals, respectively), while the other Ni$_S$ cation is at the center of a compressed O$_6$ octahedra and has no magnetic moment for the two functionals. This is compatible with the disproportionation of Ni$^{3+}$ cations to Ni$_L^{2+}$ ($t_{2g}^6 e_g^2$) and Ni$_S^{4+}$ ($t_{2g}^6 e_g^0$) cations, which are located in large (Ni$_L$) and compressed (Ni$_S$) octahedra, respectively. This results in a rock-salt pattern of 2+/4+ cations (see Figure 2.c of Ref.[8] for a sketch of the lattice distortion). As inferred by the density of states projected on all O $p$ states as well as on Ni$_L$ and Ni$_S$ $d$ states reported in Fig 1, both SCAN and PBEsol+U functionals result in an insulating ground state with a band gap of 0.90 eV and 0.54 eV, respectively, formed between a "split-off" occupied band of mainly hybridized Ni$_L^{2+}$ $e_g$ and O $p$ states and unoccupied Ni$_S^{4+}$ $e_g$ states and O $p$ states. The situation is consistent with Resonant Inelastic X-ray Spectroscopy measurements in NdNiO$_3$[12] showing the existence of ligand holes in the materials – *i.e.* unoccupied O $p$ states. It follows that hole doping might preferably – if possible – localize (form) a polaron (an acceptor state) with a Ni$_L^{2+}$ character, while electron doping might preferably localize (form) a polaron (a donor state) having a Ni$_S^{4+}$ character.

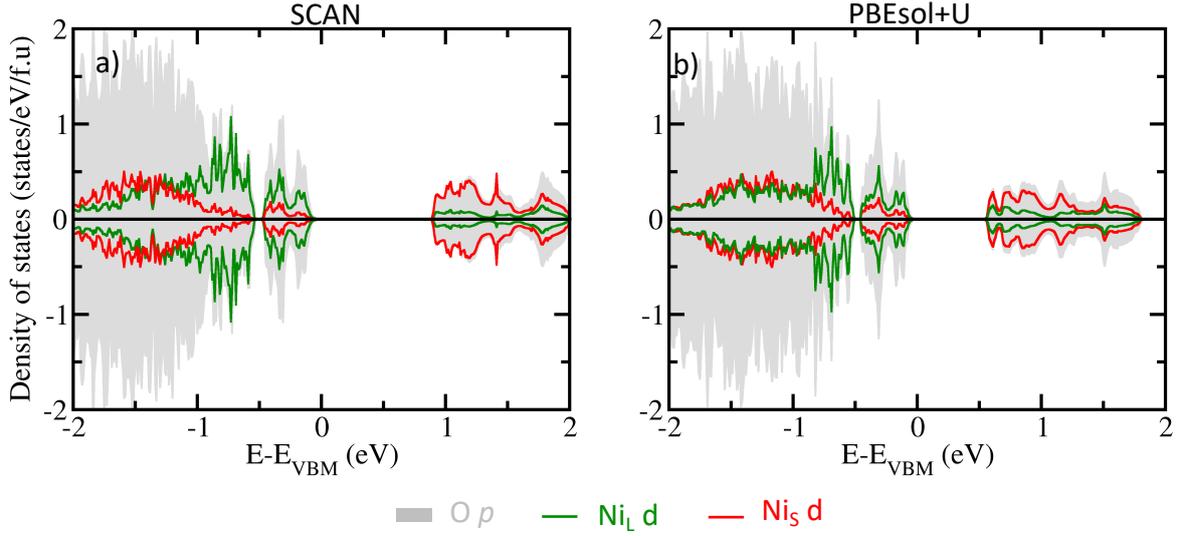

*Figure 1: Projected density of states (in states/eV/formula unit) on $Ni_L$ d (green), $Ni_S$ d (red) and O p (filled grey) states in the ground state structure of bulk $SmNiO_3$ using the SCAN (panel a) and PBEsol+U (panel b) exchange correlation functional. The 0 energy is set at the valence band maximum.*

### B. *Ability to form polarons and localized states in the band gap*

We now turn our attention to polaron formation upon hole and electron doping in $SmNiO_3$. We have first substituted one $Sm^{3+}$ cation by a $Ca^{2+}$ or $Ce^{4+}$ cation, yielding a hole or electron doping level of 6.25%, respectively. The extra hole or electron is then nudged to a precise Ni cation, located in the vicinity of the Ca/Ce introduced in the material, by forcing orbital occupancies. Nudging on both $Ni_S$ or $Ni_L$ is tested either for hole or electron doping. The results are presented only for SCAN for clarity but PBEsol+U provided the very same results – *modulo* the band gap amplitude that is smaller as in the bulk. We find that a polaron is stabilized in our ground states producing a localized intermediate state in the band gap located at $\delta E=0.10$ eV above the top of the valence band (*i.e.* an acceptor state) for hole doping or a localized intermediate state at $\delta E=0.14$ eV below the bottom of the conduction band for electron doping, as shown in Fig 2.a. and b. At the same time, the material still develops a rather large gap $\Delta E$ of 0.60 eV and 0.66 eV between the top of valence and the bottom of conduction band for hole and electron doping, respectively. As one may have anticipated, a polaron for hole (electron) doping is formed only on $Ni_L^{2+}$ ($Ni_S^{4+}$) sites in our DFT simulations. An initially nudged hole (electron) on a $Ni_S$ ($Ni_L$) cation is not stable and the solution goes back to a band theory solution with no polaron stabilized.

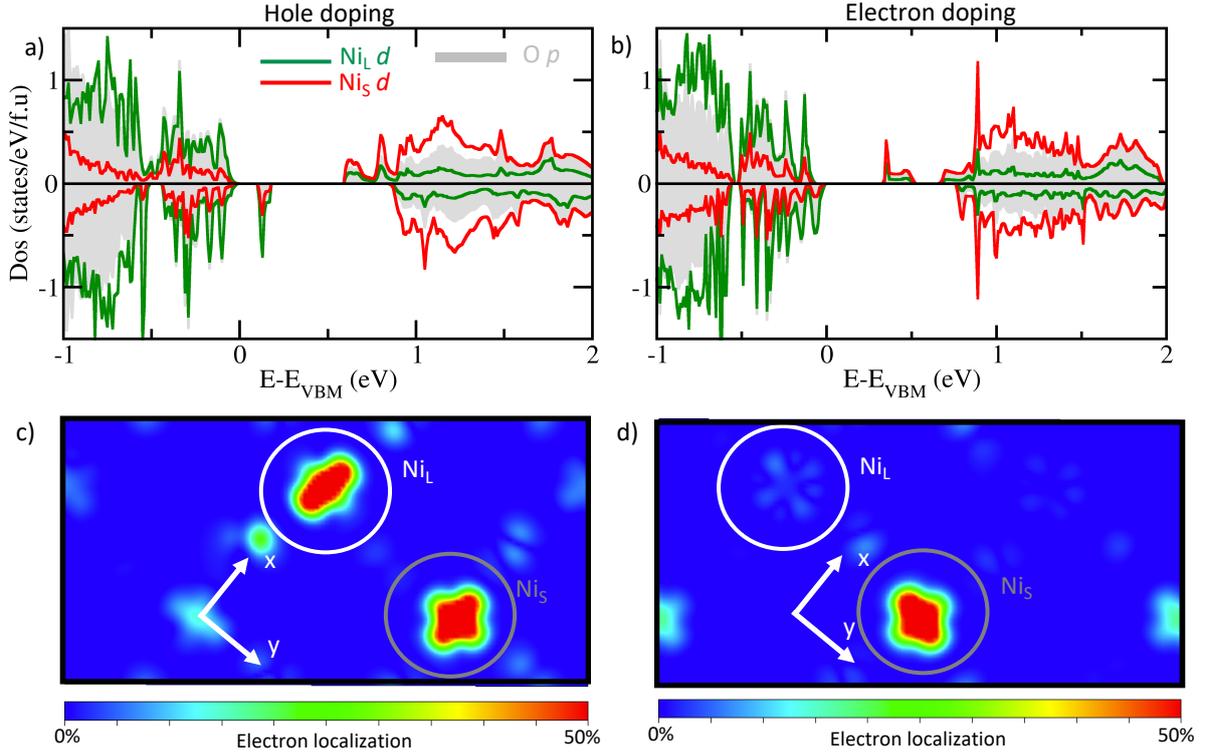

*Figure 2: a) and b) projected density of states on on $Ni_L$ d (green), $Ni_S$ d (red) and O p (filled grey) states in hole (a) and electron (b) doped $SmNiO_3$ using SCAN functional. c) and d) magnetic density difference between 6.25% hole (c) and electron (d) doped and pristine $SmNiO_3$ with the SCAN functional. The 0 energy is set at the valence band maximum.*

By looking at the charge density difference between doped and undoped material, we cannot appreciate any noticeable difference on the electronic structure of Ni sites. This is confirmed by the sphere integrated charge difference around the different Ni sites, that reveals no significant deviation of the total Ni *d* occupancies between doped and undoped materials – the largest difference of 0.08 electron is reached for the $Ni_L$ or $Ni_S$ site holding the polaron. This fact comes from a charge self-regulation mechanism[40] in which the electronic density around transition metals reorganizes itself in order to preserve the total number of *d* electrons on the TM (*i.e.* 8 electrons for Ni cations).

Nevertheless, one can extract subtle information on the electronic structure modification upon doping using magnetic moments[41]. Through hole (electron) doping, the $Ni_L$ ($Ni_S$) cation holding the polaron now exhibits a magnetic moment of 0.87/0.63 $\mu_B$ (1.04/0.78 $\mu_B$) with SCAN/PBEsol+U functional, compatible with a 3+ formal oxidation state, *i.e.* an a non-disproportionated state. At the same time, the modification of Ni magnetic moments induces an asymmetry of the spin-polarized potential experienced by each $Ni_S$ and $Ni_L$ elements. It results in small variations of the surrounding spins amplitude, either on $Ni_S$ or $Ni_L$ cation, with the magnetic moment of Ni$_S$ cations not necessarily being exactly zero. Looking at the magnetic density difference between the doped and pristine material – at fixed structure to the doped material – from Fig 2.c and d, we identify that (i) the polaron has an $e_g$ symmetry on the Ni$_L$ (Ni$_S$) site, with a $d_x^2$ ($d_y^2$) character for hole (electron) doping, and (ii) there are contributions on surrounding Ni$_S$ sites for the hole doping (Fig 2c). The former result is accompanied by a Jahn-Teller like motion that accommodates the "orbital shape" of the spin-polarized singly occupied "$e_g$-like" state, with an asymmetry of Ni$_L$-O (Ni$_S$-O) bond length in the (ab)-plane of 2.00/1.92 Å for hole (electron) doping, while the latter observation comes from the asymmetry of the spin-dependent potential.

### C. *Trends in "heavy" hole and electron doping*

Having established the ability of polaron formation, we have next performed a systematic study of electronic and structural properties upon hole and electron doping up to 25% by using both PBEsol+U and SCAN functionals. Because disproportionation effect are responsible for the metal to insulator transition (MIT) in rare-earth nickelates[6,7,11,13], we first focus on the asymmetry of magnetic moments between Ni$_L$ and Ni$_S$ sites – associated with the formal oxidation state disproportionation in the bulk MIT – and on the amplitude of the breathing mode distortion B$_{oc}$ – associated with the resulting bond disproportionation. Trends upon doping are reported in Figure 3. Again, one can see that PBEsol+U and SCAN provide very similar results for the electronic structure upon doping. The asymmetry of magnetic moment between the two Ni cations existing in the pristine compound diminishes upon hole and electron doping, indicating a loss of charge disproportionation (Fig 3.a and b) for both functionals. This is confirmed by the dependence of the breathing mode B$_{oc}$ amplitude upon doping reported in Fig3.c and d that follows a similar trend. Nevertheless, an unexpected feature for the electron doping should be highlighted: while the disproportionation totally vanishes at 25% of hole doping, it seems to strengthen again at 25% for electron doping. However, this artefact arises because the bulk "up-0-down-0" spin chains are suppressed by "heavy" electron doping, whereas the "up-0-down-0" spin chains are less affected by the presence of hole doping. More complex magnetic solutions such as paramagnetism may be required to study such heavy doping.

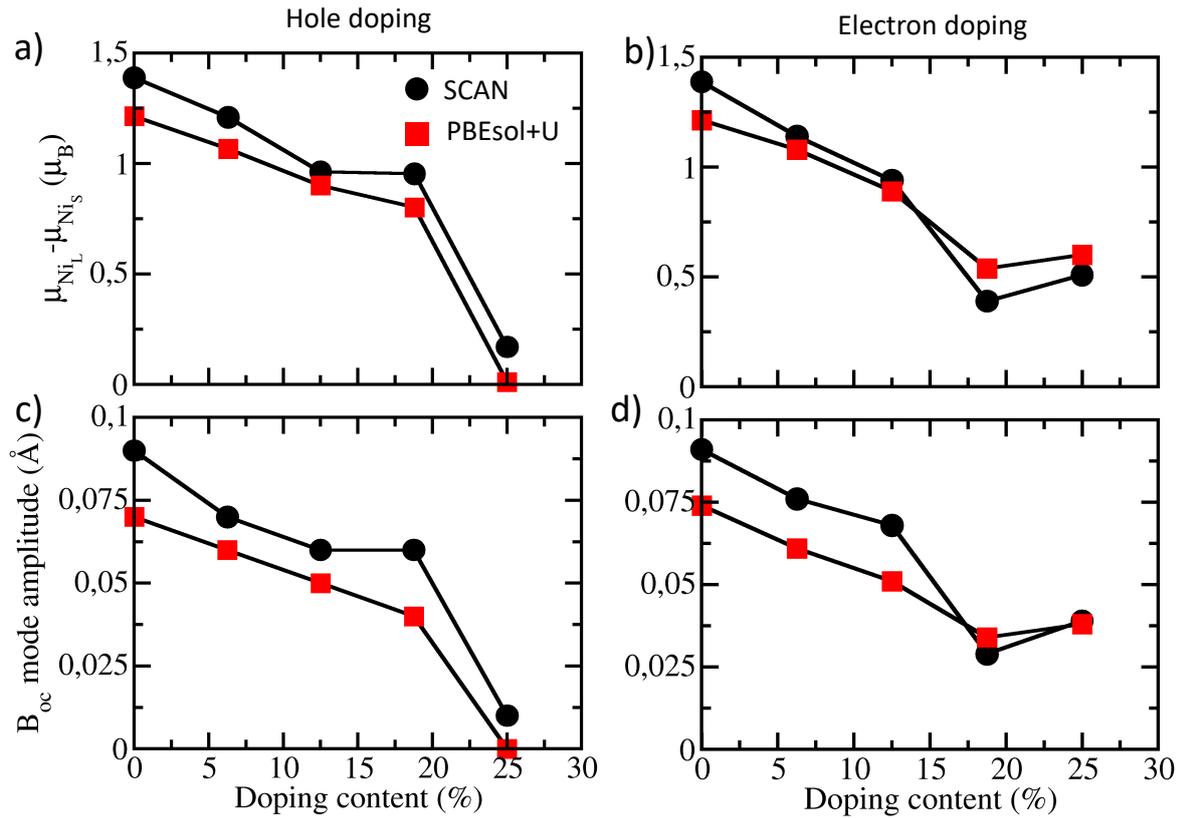

*Figure 3: Trends in (a) magnetic moment asymmetry of Ni cations (in μB) and (b) breathing distortion (in A°) as a function of the doping content (in %).*

To gain insight into the electronic properties, we report the density of states projected on O $p$ as well as on $Ni_L$ and $Ni_S$ $d$ states for increasing doping levels for the SCAN functional, cf. Figure 4. We observe that the intermediate localized states spread more and more inside the gap as the substitution of cation A increases until they merge with the initial valence/conduction band, ultimately resulting in metallic states at 25% of Sm site substitutions with Ca or Ce. Nevertheless, there is a slight difference between hole and electron doping: hole doping likely results in a semiconducting state at 18.75% with a band gap of 0.15 eV, while electron doping already yields a metal. Similar results are observed with the PBEsol+U functional although the gap at 18.75% of hole doping is reduced to 0.08 eV.

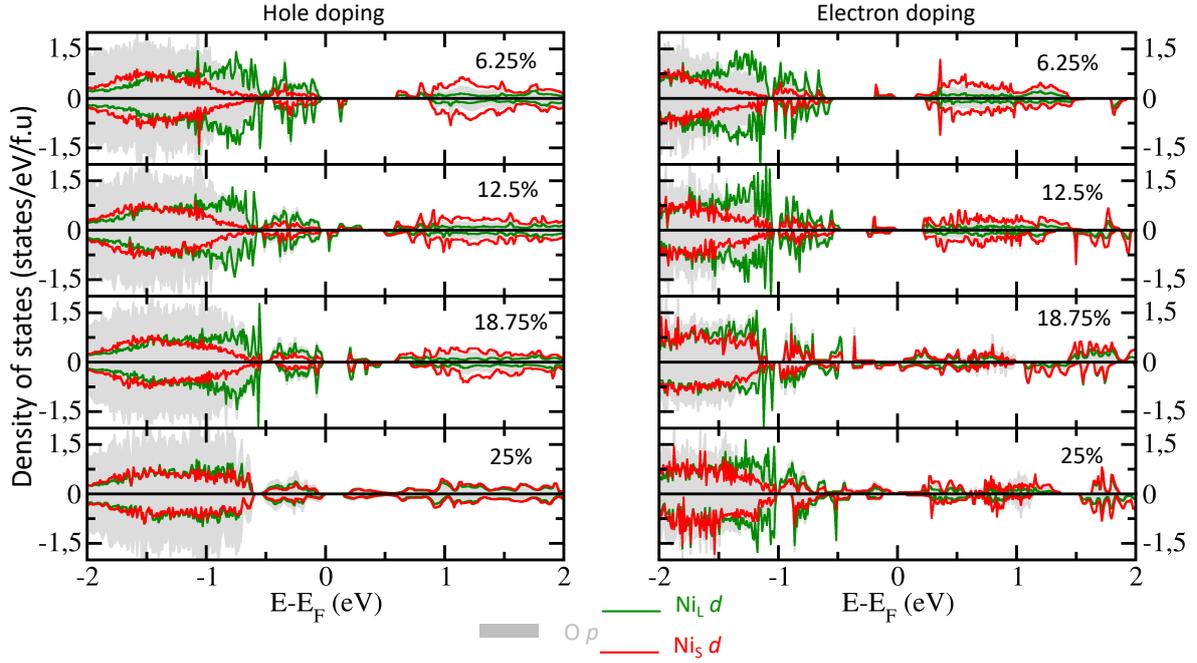

*Figure 4: Projected density of states (states/eV/f.u) on O p states (filled grey), $Ni_L$ d (green line) and $Ni_S$ d (red line) as a function of hole (left panel) and electron (right) doping content. The 0 energy is set to the Fermi Level $E_F$ at 0K, thus at the top of conduction band for hole doping and top of donor states for electron doping.*

## IV. Discussion

At low doping content (6.25%), DFT simulations performed with PBEsol+U and SCAN functional predicts an insulating character, in agreement with experimental results for both hole and electron doping in $RNiO_3$[23–25]. At larger doping content, we observe a gap closure with an asymmetry for electron and hole doping content, again compatible with experimental results of Garcia-Muñoz *et al*[23] and Xiang *et al*[24]. Nevertheless, these experimental works reveal (i) a gap closure at lower doping and that (ii) electron doping has a weaker effect on the closure of the MIT than hole doping. Our DFT work predicts a slightly different picture with a higher critical value for doping content before reaching the metallic state (up to 18.75%), and a higher hole carrier concentration with respect to electron carrier concentration before reaching a metal. These discrepancies might have several origin: (i) Xiang *et al*[24] studied Ca doping in $SmNiO_3$ but in thin films, and we cannot rule out the possibility that strain affects the metal-insulator transition upon doping[42], (ii) Garcia-Muñoz *et al*[23] examined bulk materials but with alternative cations (Nd instead of Sm, Th instead of Ce) and steric effects may alter the critical values in the spirit of the MIT temperature in bulk $RNiO_3$, (iii) we have not modelled paramagnetism in our DFT simulations, although we expect that the low temperature, spin-ordered, phase inherits the physics of the PM phase[7,28] and (iv) one might use larger supercells in our DFT calculations in order to represent more possible local motif for Ni cations.

Finally, the breathing mode distortion $B_{oc}$ is intrinsically connected to an electronic instability of the Ni +3 formal oxidation state in the bulk nickelates. This can be observed by plotting the potential energy surface as a function of the breathing mode $B_{oc}$ starting from a high symmetry cubic cell with the AFM-S order. In the undoped material, we observe a single well potential whose energy minimum is located at non-zero amplitude of $B_{oc}$. This indicates an electronic instability yielding the disproportionation effects, with a $Ni_L$ ion bearing a magnetic moment larger than one, while the $Ni_S$ cations bears no magnetic moment (see Fig 5, filled black circles). We have then checked the precise role of electron and hole doping in $SmNiO_3$ on this electronic instability by recomputing the potential energy surface. Unfortunately, due to the absence of octahedra rotations that usually help localizing electrons in perovskites, we could not stabilize solutions for doping contents higher than 6.25% with SCAN functionals. We can nevertheless observe that upon hole or electron doping (Fig.5), the energy gain associated with the $B_{oc}$ mode decreases with respect to the pristine material and the minimum is now located to lower values of $B_{oc}$ for hole doping. Thus, it confirms that doping tends to progressively remove the electronic instability associated with the MIT – charge and bond disproportionation effects – and that from pure electronic effects, hole doping suppresses the disproportionation instability – and thus the MIT – more rapidly than electron doping. This is in agreement with the indirect experimental determination of the intrinsic electronic contribution to the MIT of Garcia-Muñoz *et al*[23] – authors artificially removed the steric effect contribution to the MIT.

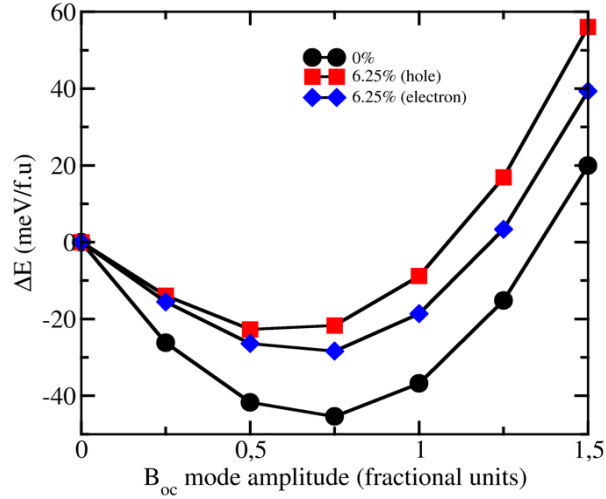

*Figure 5: Potential energy surface (in meV/f.u) as a function of a breathing mode amplitude (in fractional units) condensed in a high symmetry cubic cell as function of doping content. An amplitude of 1 corresponds to the amplitude appearing in the pristine material ground state.*

## V.     Conclusion

In conclusion, using DFT calculations, we simulated the effect of hole and electron doping in bulk $SmNiO_3$ by substituting of $Sm^{3+}$ by $Ca^{2+}$ and $Ce^{4+}$, respectively, and studied the polaron formation. In particular, we showed that DFT, even without any U parameter but with a functional amending the self-error interaction inherent to DFT, produces a polaronic state in the band gap at low doping. This results in acceptor (hole doping) or donor (electron doping) states and a semi-conducting behavior at moderate doping concentrations. As both hole and electron doping increases, the gap decreases and the bond disproportionation vanishes, until reaching a metallic state at large doping concentrations. Our results are globally in line with experimental reports for both bulk and thin films, which validates the use of DFT and of the meta-GGA SCAN functional in capturing doping effects in $RNiO_3$ materials. Therefore, these results open the way to study these compounds in form of thin films by using DFT without U parameter and even extend the study to superconducting infinite-layer nickelates.

## Acknowledgements

We acknowledge access granted to HPC ressources of Criann through the projects 2020005 and 2007013 and of Cines through the DARI project A0080911453). Support from the French ANR project "QUANTOP" is acknowledged.